\documentclass[12pt, draftclsnofoot, onecolumn]{IEEEtran}
\IEEEoverridecommandlockouts
\usepackage{epsfig,graphicx,subfigure,psfrag,amsmath,cases,bm}
\usepackage{latexsym,amssymb,algorithm,mathtools}
\usepackage{algorithmic}
\usepackage{color}
\usepackage{url}
\usepackage{scrtime}
\usepackage{stfloats}
\usepackage{tablefootnote}
\usepackage{cite}
\usepackage{amsfonts,mathabx}
\usepackage{textcomp}
\usepackage{xcolor,capt-of}
\usepackage{multirow}
\usepackage{amsthm}
\makeatletter
\newcommand*{\rom}[1]{\expandafter\@slowromancap\romannumeral #1@}
\makeatother
\DeclareMathOperator{\maxo}{maximize}

\newtheorem{remark}{Remark}

\newtheorem{assumption}{Assumption}
\newtheorem{theorem}{Theorem}
\newcommand{\QED}{\hfill \ensuremath{\blacksquare}}
\allowdisplaybreaks
\newtheorem{assumptionalt}{Assumption}[assumption]

\newenvironment{assumptionp}[1]{
  \renewcommand\theassumptionalt{#1}
  \assumptionalt
}{\endassumptionalt}

\title{Active IRS-Aided MIMO Systems: How Much Gain Can We Get?}
\author{\IEEEauthorblockN {Zeyan Zhuang, Xin Zhang, Dongfang Xu, and Shenghui Song}

Dept. of ECE, The Hong Kong University of Science and Technology, Hong Kong\\
}
\begin{document}
\maketitle

\begin{abstract}
Intelligent reflecting surfaces (IRSs) have emerged as a promising technology to improve the efficiency of wireless communication systems. However, passive IRSs suffer from the  ``multiplicative fading" effect, because the transmit signal will go through two fading hops. With the ability to amplify and reflect signals, active IRSs offer a potential way to tackle this issue, where the amplification energy only experiences the second hop. However, the  fundamental limit and system design for active IRSs have not been fully understood, especially for multiple-input multiple-output (MIMO) systems. In this work, we consider the analysis and design for the large-scale active IRS-aided MIMO system assuming only statistical channel state information (CSI) at the transmitter and the IRS. The evaluation of the fundamental limit, i.e., ergodic rate, turns out to be a very difficult problem. To this end, we leverage random matrix theory (RMT) to derive the deterministic approximation (DA) for the ergodic rate, and then design an algorithm to jointly optimize the transmit covariance matrix at the transmitter and the reflection matrix at the active IRS. Numerical results demonstrate the accuracy of the derived DA and the effectiveness of the proposed optimization algorithm. The results in this work reveal interesting physical insights with respect to the advantage of active IRSs over their passive counterparts.
\end{abstract}

% \begin{IEEEkeywords}
% component, formatting, style, styling, insert
% \end{IEEEkeywords}

\section{Introduction}
With the development of innovative applications, there is an increasing demand for higher data rate, reliability, and energy efficiency in future wireless communication systems. To this end, intelligent reflecting surfaces (IRSs), have been proposed as an energy-efficient way to create a favorable channel between the transmitter and receiver \cite{wuqingqing2019, Huang2019}. Specifically, IRSs can reshape the wireless channel and improve the signal quality by reflecting signals through a surface composed of a large number of reconfigurable elements. 
\par
Many engaging results have been obtained on the analysis and design of IRS-aided systems such as the IRS-aided multiple-input single-output (MISO) \cite{wuqingqing2019} and multiple-input multiple-output (MIMO) channels \cite{Zhang2020, Jun2020, Xin2022}. In particular, \cite{Zhang2020} studied the fundamental limit of IRS-aided point-to-point MIMO systems with perfect channel state information (CSI) at the transmitter, receiver, and IRS. However, in practice, perfect CSI is extremely difficult to obtain for IRS-aided systems, especially in the case of fast-fading channels \cite{Nadeem2021}. As a result, \cite{Jun2020} investigated the achievable rate of a large-scale passive IRS-aided MIMO system where only statistical CSI is available at the transmitter and the IRS, by random matrix theory (RMT). The design of the transmit covariance matrix and the phase-shift matrix was also considered. 
\par
With either perfect or statistical CSI, the potential gain brought by IRSs is often limited by the ``multiplicative fading"  of the two-hop channel \cite{Najafi2020}. To circumvent this issue, a new IRS architecture, namely, active IRS, has recently been proposed in \cite{Zijian2023}. In particular, equipped with the reflection-type amplifiers (e.g., current-inverting converters), active IRSs can not only reflect the signal to a desired direction, but also amplify the signal with additional power \cite{You2021}. It is worth mentioning that active IRSs and full-duplex (FD) amplify-and-forward (AF) relays differ in both hardware implementation and transmission modes \cite{wuqingqing2019}. Specifically, FD AF relays lead to unavoidable self-interference and processing delay when amplifying the received signal. In contrast, active IRSs can amplify signals instantaneously without introducing self-interference, although the processing freedom is limited, i.e., the diagonal reflection matrix. Inspired by these advantages, some researchers have investigated the analysis and design of active IRS-aided systems. In particular, \cite{Zhi2022letter, You2021} compared the performance of active and passive IRS-aided single-input single-output (SISO) systems given the same overall power budget. The authors of \cite{zhou2023framework} investigated the fundamental limits of IRS-aided MISO systems with partial CSI. However, to the best of the authors' knowledge, the fundamental limits of active IRS-aided MIMO systems are not yet available in the literature. 
\par
Motivated by the above works, in this paper, we investigate the active IRS-aided point-to-point MIMO communication system assuming only statistical CSI at the transmitter and the IRS. The objective is to first determine the fundamental limit of the concerned system, i.e., the ergodic rate. Then, based on the evaluation result, we jointly design the transmit covariance matrix and the reflection matrix to maximize the ergodic rate. We achieve the first objective by deriving the deterministic approximation (DA) of the ergodic rate using RMT. As the optimization problem is non-convex, we propose an alternating optimization (AO)-based algorithm to design the optimal transmit covariance matrix and reflection matrix. Numerical results validate the accuracy of the DA and effectiveness of the proposed algorithm. 
\par
The main contributions of this work include:
\begin{itemize}
\item We evaluate the ergodic rate of active IRS-aided MIMO systems with only statistical CSI at the transmitter and the IRS. To obtain a more tractable and computationally-efficient form for ease of use in the following optimization, we derive the DA for the ergodic rate.
\item We maximize the DA for the ergodic rate by jointly optimizing the transmit covariance matrix and reflection matrix. To tackle the resulting non-convex optimization problem, an AO-based low-complexity suboptimal algorithm is developed.
% \item We investigate the impact of the dynamic noise introduced by active IRS and compare the active IRS and the passive one under different settings. Besides, our simulation results show that active IRS can bring gain even without system design and for a small number of elements, active IRS can significantly improve the rate gain compared to passive IRS. While as the size of the IRS becomes larger, the aforementioned performance gap gradually vanishes.
\item  We investigate the impact of the dynamic noise introduced by the active IRS and different power allocation policies. Simulation results unveil that the active IRS is effective to tackle the ``multiplicative fading” effect and has consistent advantages over the passive one.
%reveal that under which condition active IRS perform better.  and the proposed AO-based algorithm is effective. By design the transmit covariance matrix of transmitter and the reflection matrix of the active IRS, significant rate gain can be achieved.
\end{itemize}
%%%%%%%%%%%%%%%%%%%%%%%%%%%%%%%%%%%%%%
% \begin{figure*}[t]
% 	\setcounter{equation}{1}
% \begin{equation}\label{rate}
%      R(\mathbf{Q}, \mathbf{\Phi}) = \frac{1}{n_r}\mathbb{E} \Big[\log \det \big(\frac{\mathbf{H}_1 \mathbf{\Phi} \mathbf{H}_2 \mathbf{Q} \mathbf{H}_2^H \mathbf{\Phi}^H \mathbf{H}_1^H}{\sigma_s^2}
%     + \frac{\sigma_d^2\mathbf{H}_1\mathbf{\Phi}\mathbf{\Phi}^H\mathbf{H}_1^H}{\sigma_s^2} + \mathbf{I} \big)
%     - \log \det\big( \frac{\sigma_d^2\mathbf{H}_1\mathbf{\Phi}\mathbf{\Phi}^H\mathbf{H}_1^H}{\sigma_s^2} + \mathbf{I} \big)\Big], 
% \end{equation}
% 			\hrule
% \end{figure*}
% \setcounter{equation}{0}
%%%%%%%%%%%%%%%%%%%%%%%%%%%%%%%%%%%%%
\par
The rest of this paper is organized as follows. In section \ref{S2}, we introduce the system model and formulate the problem. In section  \ref{S3}, we derive the DA for the ergodic rate. An AO-based algorithm for optimizing the transmit covariance matrix and the reflection matrix is proposed in section \ref{S4}. Section \ref{S5} gives the numerical results and
Section \ref{S6} concludes the paper. The notations utilized in this paper are listed in the footnote\footnote{\underline{Notations}. In this paper, we use boldfaced uppercase letters and lowercase letters to represent matrices and column vectors, respectively. $\Im(a)$ and $\Re(a)$ denote the imaginary and real part of a complex number $a$, respectively. $\mathbb{R}^+$ and $\mathbb{C}^+$ denote the real non-negative axis $\{x \in \mathbb{R}: x \geq 0\}$ and the upper half plane $\{z \in \mathbb{C}: \Im(z) > 0\}$, respectively. Let $\jmath 
 = \sqrt{-1}$. The $(i, j)$-th element of matrix $\mathbf{A}$ will be denoted as either $a_{ij}$ or $[\mathbf{A}]_{ij}$. $\mathrm{diag}\{\mathbf{a}\}$ denotes the diagonal square matrix whose diagonal entries are elements of vector $\mathbf{a}$ and $\mathrm{diag}\{\mathbf{A}\} = \mathrm{diag}\{a_{ii}; 1 \leq i \leq n\}$ for all $n \times n$ matrix $\mathbf{A}$. The superscript `$H$' denotes Hermitian transpose operator. $\mathrm{Tr}(\mathbf{A})$ and $||\mathbf{A}||$  represent the trace and spectral norm of $\mathbf{A}$, respectively.  $\mathbb{E}[a]$ denotes the expectation of random variable $a$. Function $(a)^+ = \max\{a, 0\}$. $\xrightarrow[\mathrm{X}]{a.s.}$ represents the almost sure convergence under the process X. $\mathrm{supp}(\mu)$ denotes the support of measure $\mu$.}. 
\begin{figure}[t]
\centering
% \subfigure[Normalized ergodic rate versus static noise power.]{
%     \centering
%     \includegraphics[width=3.2in]{figure_1_1.eps}
% }
\includegraphics[width=3.8in]{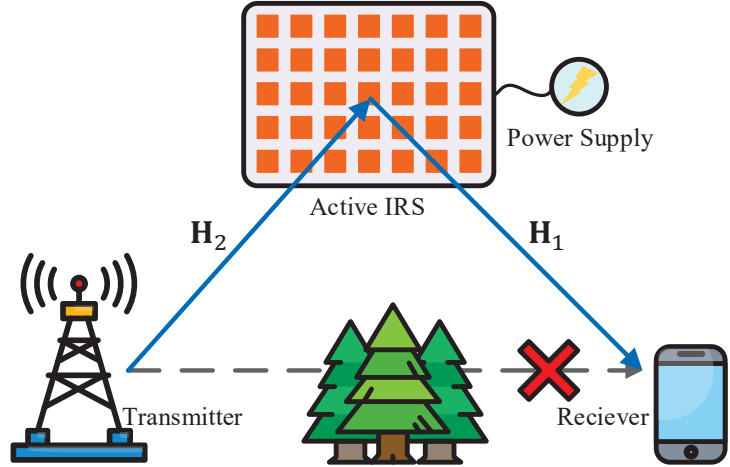}

% \subfigure[Normalized ergodic rate versus dynamic noise power.]{
%     \centering
%     \includegraphics[width=3.2in]{figure_1_2.eps}
% }
\caption{Active IRS-aided communication system.}\label{System_model}
\end{figure}
%%%%%%%%%%%%%%%%%%%%%%%%%%%%%%%%%%%%%%%%%%%%%%%%%%%
\section{System model and Problem Formulation} \label{S2}
\subsection{System Model}
As shown in Fig. \ref{System_model}, we consider a MIMO system consisting of a base station (BS) equipped with $n_t$ transmit antennas and a user equipped with $n_r$ receive antennas. Due to the blockage of the direct link, an IRS comprising $n_l$ active reflecting elements is utilized to establish an alternative communication link. Due to the active nature, the thermal noise introduced during amplification cannot be ignored \cite{Zijian2023}. Therefore, the received signal $\mathbf{y}$ at the user side is given by
\begin{align}
 \mathbf{y} = \mathbf{H}_1 \mathbf{ \Phi} \mathbf{H}_2 \mathbf{x} + \mathbf{H}_1 \mathbf{\Phi} \mathbf{n}_d+ \mathbf{n}_s, 
\end{align}
where $ \mathbf{x} \in \mathbb{C}^{n_t}$ denotes the transmit signal with covariance matrix $ \mathbf{Q} = \mathbb{E}[\mathbf{x} \mathbf{x}^H]$. The matrices $ \mathbf{ H}_2 \in \mathbb{C}^{n_l \times n_t}$, $ \mathbf{ H}_1 \in \mathbb{C}^{n_r \times n_l}$ represent BS-IRS and IRS-user channel, respectively. The reflection matrix of the active IRS is denoted by $\mathbf{ \Phi} = \mathrm{diag} \{a_1 \phi_1,a_2 \phi_2, \cdots ,a_{n_l} \phi_{n_l}\}$, where $a_i \in \mathbb{R^+}$ and $\phi_{i} = e^{\jmath\theta_i}$, $i=1, 2, \cdots, n_l$ represent the amplification factor and phase shift of the $i$-th IRS element, respectively. Different from the passive IRS, $a_i$ can be larger than $1$. Here $\mathbf{n}_d \in \mathbb{C}^{n_l}$ denotes the dynamic noise introduced by the active component and $\mathbf{n}_s \in \mathbb{C}^{n_r}$ is the static noise at the receiver. Both $\mathbf{n}_d$ and $\mathbf{n}_s$ are modeled as additive Gaussian white noise (AWGN)\cite{Zijian2023} \cite{Dongfang2021}, i.e.,
$\mathbf{n}_d \sim \mathcal{CN}(\mathbf{0}, \sigma_d^2 \mathbf{I})$, $\mathbf{n}_s \sim \mathcal{CN}(\mathbf{0}, \sigma_s^2 \mathbf{I})$.
\par
Assuming only statistical CSI at the transmitter and the IRS, the normalized ergodic rate of the channel is given by 
\begin{eqnarray}
     R(\mathbf{Q}, \mathbf{\Phi}) \hspace{-2mm} &=&\hspace{-2mm} \frac{1}{n_r}\mathbb{E} \Big[\log \det \big( \frac{1}{\sigma_s^2}\mathbf{H}_1 \mathbf{\Phi} \mathbf{H}_2 \mathbf{Q} \mathbf{H}_2^H \mathbf{\Phi}^H \mathbf{H}_1^H + \frac{\sigma_d^2}{\sigma_s^2}\mathbf{H}_1\mathbf{\Phi}\mathbf{\Phi}^H\mathbf{H}_1^H + \mathbf{I} \big) \notag\\
    \hspace{-2mm} &-&\hspace{-2mm}  \log \det\big( \frac{\sigma_d^2}{\sigma_s^2}\mathbf{H}_1\mathbf{\Phi}\mathbf{\Phi}^H\mathbf{H}_1^H + \mathbf{I} \big)\Big] \label{rate}, 
\end{eqnarray}
where $R(\mathbf{Q}, \mathbf{\Phi})$ has the unit of bits per second per Hz per antenna. The channel matrix can be written by the Kronecker model as 
\setcounter{equation}{2}
\begin{align}
\mathbf{H}_1 = \mathbf{R}_1^{\frac{1}{2}} \mathbf{X}_1 \mathbf{T}_1^{\frac{1}{2}} \mbox{ and }  \mathbf{H}_2 =  \mathbf{R}_2^{\frac{1}{2}} \mathbf{X}_2 \mathbf{T}_2^{\frac{1}{2}}, 
\end{align}
where $\mathbf{R}_1 \in \mathbb{C}^{n_r \times n_r}$, $\mathbf{T}_1 \in \mathbb{C}^{n_l \times n_l}$ ,
$\mathbf{R}_2 \in \mathbb{C}^{n_l \times n_l}$ and $\mathbf{T}_2 \in \mathbb{C}^{n_t \times n_t}$ are Hermitian non-negative definite matrices. $\mathbf{R}_1$ and $\mathbf{T}_2$ represent the spatial correlation matrices for the receiver and the transmitter, while $\mathbf{R}_2$ and $\mathbf{T}_1$ denote the spatial correlation matrices for the IRS. We utilize $L_1$ and $L_2$ to denote the path loss for the IRS-user and BS-IRS link, respectively. For ease of manipulation, $L_i$ is absorbed into $\mathbf{R}_i$. $\mathbf{X}_1 = \mathbb{C}^{n_r \times n_l}$ and $\mathbf{X}_2 \in  \mathbb{C}^{n_l \times n_t}$ are matrices whose entries are independent and identically distributed (i.i.d.) Gaussian random variables (r.v.s) with $[\mathbf{X}_1]_{ij} \sim \mathcal{CN}(0, \frac{1}{n_l})$, $[\mathbf{X}_2]_{ij} \sim \mathcal{CN}(0, \frac{1}{n_t})$. For simplicity, we define
\begin{equation}
\widetilde{\mathbf{T}}_1 = \mathbf{\Phi}^H \mathbf{T}_1 \mathbf{\Phi} \mbox{ and }\widetilde{\mathbf{T}}_2 = \mathbf{Q}^{\frac{1}{2}} \mathbf{T}_2 \mathbf{Q}^{\frac{1}{2}}.
\end{equation}
Due to the rotational invariance of Gaussian distribution, the normalized ergodic rate can be written as
\begin{equation}
    R(\mathbf{Q}, \mathbf{\Phi}) = \frac{1}{n_r}\mathbb{E}\big[\log \det \big(\frac{\mathbf{B}_1}{\sigma_s^2} + \mathbf{I} \big) - \log \det\big( \frac{\mathbf{B}_2}{\sigma_s^2} + \mathbf{I} \big)\big], \label{Rate_form}
\end{equation}
where $\mathbf{B}_1 = (\widetilde{\mathbf{H}}_1  \widetilde{\mathbf{H}}_2  \widetilde{\mathbf{H}}_2^H  \widetilde{\mathbf{H}}_1^H + \sigma_d^2\widetilde{\mathbf{H}}_1\widetilde{\mathbf{H}}_1^H )$, $\mathbf{B}_2 =\sigma_d^2  \widetilde{\mathbf{H}}_1\widetilde{\mathbf{H}}_1^H$, $\widetilde{\mathbf{H}}_1 =  \mathbf{R}_1^{\frac{1}{2}} \mathbf{X}_1 \widetilde{\mathbf{T}}_1^{\frac{1}{2}}$, and $\widetilde{\mathbf{H}}_2 = \mathbf{R}_2^{\frac{1}{2}} \mathbf{X}_2 \widetilde{\mathbf{T}}_2^{\frac{1}{2}}$.
\subsection{Problem Formulation}
Our objective is to optimize the normalized ergodic rate of the active IRS-aided MIMO system under the power constraint at both the transmitter and IRS. The problem can be formulated as 
\begin{eqnarray}
\label{Problem1}
    \mathcal{P}_1:&&\underset{\mathbf{Q}\succeq\mathbf{0},\bm{\Phi}}{\maxo}\hspace*{2mm}R(\mathbf{Q},\bm{\Phi})\notag\\
    &&\hspace*{5mm}\mbox{s.t.}\hspace*{7mm}\mbox{C1:}\hspace*{1mm}\mathrm{Tr} (\mathbf{Q}) \leq n_t P_{T},\notag\\
    &&\hspace*{17mm}\mbox{C2:}\hspace*{1mm}\frac{\mathrm{Tr}(\mathbf{Q} \mathbf{T}_2)}{n_t}\mathrm{Tr}(\mathbf{R}_2(\mathbf{\Phi}\mathbf{\Phi}^H  - \mathbf{I})) + \sigma_d^2 \mathrm{Tr}(\mathbf{\Phi}\mathbf{\Phi}^H)\leq P_{A},
\end{eqnarray}
where C1 represents the maximum power constraint at the transmitter and C2 denotes the amplification power budget of the active IRS. 
\begin{remark}
The received signal of the active IRS is $\mathbf{s}_{in} = \mathbf{H}_2 \mathbf{x}$ and the reflected signal is $\mathbf{s}_{out} = \mathbf{ \Phi} \mathbf{H}_2 \mathbf{x} + \mathbf{\Phi} \mathbf{n}_d$. The amplification
power of the active IRS is $\mathbb{E}[\mathbf{s}_{out}^H \mathbf{s}_{out}]$ \cite{Zijian2023, wuqingqing2019}. However, if we set the power constraint as $\mathbb{E}[\mathbf{s}_{out}^H \mathbf{s}_{out}] \leq P_A$, then when $P_A \rightarrow 0$, i.e., there is no power supply at IRS, the amplification factor $a_i$ will become $0$. To ensure a fair comparison with the passive IRS, we subtract the received energy of the IRS from the amplification energy and formulate the constraint as in C2. Therefore, the total power consumed by the active IRS is $\mathbb{E}[\mathbf{s}_{out}^H \mathbf{s}_{out}] - \mathbb{E}[\mathbf{s}_{in}^H \mathbf{s}_{in}] $. 
\end{remark}
\par
Note that $\mathcal{P}_1$ is very challenging to solve for two reasons. Firstly, the objective function is the expectation over a log determinant. Secondly, the objective function is the difference between two terms and the constraints C1 and C2 are coupled, making the optimization problem non-convex. In the following, we first derive the DA to approximate the normalized ergodic rate and then propose an effective AO-based algorithm to jointly optimize the transmit covariance matrix and the reflection matrix.
%%%%%%%%%%%%%%%%%%%%%%%%%%%%%%%%%%%%%%%%%%%%%%%%%%%%%%%%%%%
\section{Deterministic Approximation of Ergodic Rate} \label{S3}
Direct evaluation of the ergodic rate is very difficult. In this section, we first derive the DA for the normalized ergodic rate by leveraging RMT. For that purpose, we first introduce two important assumptions. 
\begin{assumptionp}{A-1} \label{A-1}
Let $c_1 = \frac{n_r}{n_l}$, $c_2 = \frac{n_l}{n_t}$, then $0 < \lim \inf c_1 \leq \lim \sup c_1 < +\infty$, $0 < \lim \inf c_2 \leq \lim \sup c_2 < +\infty$.
\end{assumptionp}
\begin{assumptionp}{A-2}\label{A-2}
$\underset{\overline{n} \rightarrow + \infty}{\limsup} \hspace{1mm} (\underset{i = 1, 2}{\max}\{||\mathbf{R}_i||,\hspace*{1mm}  ||\widetilde{\mathbf{T}}_i||\}) < +\infty$.
\end{assumptionp} 
\noindent\textbf{\ref{A-1}} implies that the number of transmit antennas, receive antennas, and active IRS elements grow to $+\infty$ at the same rate. In the following, we use $\overline{n} \rightarrow + \infty$ to represent this asymptotic regime. \textbf{\ref{A-2}} implies that the antenna imbalance is finite \cite{Xin2021bias}. 
\par
For a positive measure $\mu$ over $\mathbb{R}$,  its Stieltjes transform is defined as $m_{\mu}(z) = \int_{\mathbb{R}}\frac{\mu(\mathrm{d}t)}{t-z}, z\in \mathbb{C} \backslash \mathrm{supp}(\mu)$.  We denote $\mathcal{S}(\mathbb{D})$ as the set of Stieltjes transforms for positive measures over domain $\mathbb{D}$. For matrix $\mathbf{M} \in \mathbb{C}^{n \times n}$, we denote $\mathbf{Q}_{\mathbf{M}}(z) = (-z \mathbf{I} + \mathbf{M})^{-1}$ as its resolvent matrix and $m_{\mathbf{M}}(z) = \frac{1}{n} \mathrm{Tr}\mathbf{Q}_{\mathbf{M}}(z)$ represents the Stieltjes transform of its empirical spectrum distribution (ESD).  In order to evaluate \eqref{Rate_form}, we use the Shannon transform\cite{Couillet2011DE}, i.e., $R(\mathbf{Q},\bm{\Phi}) =  \int^{+\infty}_{\sigma_s^2} \mathbb{E} [m_{\mathbf{B}_2}(-t) - m_{\mathbf{B}_1}(-t)  ]\mathrm{d}t$, where $m_{\mathbf{B}_1}(-t)$ and $m_{\mathbf{B}_2}(-t) $ are the Stieltjes transforms of the ESDs for $\mathbf{B}_1$ and $\mathbf{B}_2$ defined in \eqref{Rate_form}. To approximate the ergodic rate, we first give the two DAs of $\mathbb{E} [m_{\mathbf{B}_1}(z)]$ and $\mathbb{E}[m_{\mathbf{B}_2}(z) ]$
in the following two theorems.
\begin{theorem}
With assumptions \textbf{\ref{A-1}} and \textbf{\ref{A-2}}, we have
\begin{equation}
    \mathbb{E} [m_{\mathbf{B}_1}(z)] - \frac{1}{n_r}\mathrm{Tr}(-z\mathbf{I} + (\delta_2\delta_3 + \sigma_d^2\delta_4)\mathbf{R}_1)^{-1} \xrightarrow{\overline{n} \rightarrow +\infty} 0, \label{Th1_DE}
\end{equation}
where $\delta_i$, $i=1,2,3,4$, is the unique solution of the following system of equations
\begin{eqnarray} 
\delta_1 \hspace{-3mm}&=&\hspace{-3mm} \frac{1}{n_r} \mathrm{Tr} [\mathbf{R}_1(-z\mathbf{I} + (\delta_2\delta_3 + \sigma_d^2\delta_4)\mathbf{R}_1)^{-1}], \label{fixed_point_1}\\
\delta_2 \hspace{-3mm}&=&\hspace{-3mm} \frac{1}{n_l} \mathrm{Tr} [\widetilde{\mathbf{T}}_1^{\frac{1}{2}} \mathbf{R}_2 \widetilde{\mathbf{T}}_1^{\frac{1}{2}}(\mathbf{I}\hspace{-0.5mm}+\hspace{-0.5mm} \sigma_d^2 c_1 \delta_1\widetilde{\mathbf{T}}_1 \hspace{-0.5mm}+\hspace{-0.5mm} c_1 \delta_1 \delta_3\widetilde{\mathbf{T}}_1^{\frac{1}{2}} \mathbf{R}_2 \widetilde{\mathbf{T}}_1^{\frac{1}{2}} )^{-1}], \label{fixed_point_2}  \\
\delta_3 \hspace{-3mm}&=&\hspace{-3mm} \frac{1}{n_t} \mathrm{Tr} [\widetilde{\mathbf{T}}_2 (\mathbf{I} + c_1c_2 \delta_1 \delta_2 \widetilde{\mathbf{T}}_2)^{-1}], \label{fixed_point_3}\\
\delta_4 \hspace{-3mm}&=&\hspace{-3mm} \frac{1}{n_l} \mathrm{Tr} [\widetilde{\mathbf{T}}_1(\mathbf{I} + \sigma_d^2 c_1 \delta_1\widetilde{\mathbf{T}}_1 + c_1 \delta_1 \delta_3\widetilde{\mathbf{T}}_1^{\frac{1}{2}} \mathbf{R}_2 \widetilde{\mathbf{T}}_1^{\frac{1}{2}} )^{-1}], \label{fixed_point_4}
\end{eqnarray}
 such that $\{\delta_1, \delta_1 \delta_2, \delta_1 \delta_3, \delta_1 \delta_4\} \subset \mathcal{S}(\mathbb{R}^+)$.
\end{theorem}
\textit{Proof}: Please refer to Appendix A. \QED
\begin{theorem}
With assumptions \textbf{\ref{A-1}} and \textbf{\ref{A-2}}, we have
\begin{equation}
    \mathbb{E} [m_{\mathbf{B}_2}(z)] - \frac{1}{n_r}\mathrm{Tr}(-z\mathbf{I} + \alpha_2 \mathbf{R}_1)^{-1} \xrightarrow{\overline{n} \rightarrow + \infty} 0,
\end{equation}
where $\alpha_i$, $i=1, 2$, is the unique solution of the following system of equations
\begin{eqnarray}
\alpha_1 \hspace{-3mm}&=&\hspace{-3mm} \frac{1}{n_r} \mathrm{Tr} [\mathbf{R}_1 (-z\mathbf{I} + \sigma_d^2 \alpha_2 \mathbf{R}_1)^{-1}], \label{fixed_point_2_1}\\
\alpha_2 \hspace{-3mm}&=&\hspace{-3mm} \frac{1}{n_l} \mathrm{Tr} [\widetilde{\mathbf{T}}_1 (\mathbf{I} + c_1  \sigma_d^2  \alpha_1 \widetilde{\mathbf{T}}_1)^{-1}], \label{fixed_point_2_2}
\end{eqnarray}
such that $\{\alpha_1, -\frac{\alpha_2}{z}\} \subset \mathcal{S}(\mathbb{R}^+)$.
\end{theorem}
The proof of \textbf{Theorem 2} is similar to that of \textbf{Theorem 1} and is omitted here due to limited space. With the above two results, the DA for the normalized ergodic rate is given in the following theorem.
\begin{theorem}
$R(\mathbf{Q},\bm{\Phi}) -  \overline{R}(\mathbf{Q},\bm{\Phi}) \xrightarrow{\overline{n} \rightarrow + \infty} 0$, where
\begin{eqnarray}
    \overline{R} \hspace{-2mm}&=&\hspace{-3mm} \overline{R}_1 - \overline{R}_2, \label{DA_rate_1}\\
    \overline{R}_1 \hspace{-3mm}&=&\hspace{-3mm} \frac{1}{n_r} \big[ \log \det( \mathbf{I} + \frac{\delta_2 \delta_3 + \sigma_d^2 \delta_4}{\sigma_s^2} \mathbf{R}_1) +  \log \det(\mathbf{I} + c_1 c_2 \delta_1 \delta_2 \widetilde{\mathbf{T}}_2  ) \notag\\
    \hspace{-3mm}& +&\hspace{-3mm} \log \det(\mathbf{I} + c_1\delta_1 \sigma_d^2 \widetilde{\mathbf{T}}_1 + c_1 \delta_1 \delta_3 \widetilde{\mathbf{T}}_1 ^{\frac{1}{2}} \mathbf{R}_2 \widetilde{\mathbf{T}}_1 ^{\frac{1}{2}} ) \big]  - 2 \delta_1 \delta_2 \delta_3 - \sigma_d^2 \delta_1 \delta_4,  \label{DA_rate_2} \\
    \overline{R}_2 \hspace{-3mm}&=&\hspace{-3mm} \frac{1}{n_r} \big[\log \det(\mathbf{I} + \frac{\sigma_d^2}{\sigma_s^2} \alpha_2 \mathbf{R}_1) + \log \det(\mathbf{I} +  c_1 \sigma_d^2 \alpha_1 \widetilde{\mathbf{T}}_1)\big] - \sigma_d^2 \alpha_1\alpha_2. \label{DA_rate_3}
\end{eqnarray}
\end{theorem}
\textit{Proof}: Please refer to Appendix B. \QED
\par
\begin{remark}
    If we let $\sigma_d^2 \rightarrow 0$ in the fixed point system \eqref{fixed_point_1}-\eqref{fixed_point_4} and the DA of the normalized ergodic rate \eqref{DA_rate_1}-\eqref{DA_rate_3},  $\delta_4$ will disappear in $\overline{R}$ and $\overline{R}_2 \rightarrow 0$. Moreover, if we let $a_i = 1$, the above result will degenerate to that in \cite[Theorem 1]{Jun2020} which is the DA for the normalized ergodic rate of passive IRS-aided MIMO systems.
\end{remark}
%%%%%%%%%%%%%%%%%%%%%%%%%%%%%%%%%%%%%%%%%%%%%%%%%%%%%%%%%%%%%%%%%%%%%
\section{optimization of the transmit covariance matrix and the Reflection Matrix}\label{S4}
Based on the DA derived in the last section, we develop an optimization algorithm for maximizing $\overline{R}$ in this section. In particular, we reformulate the problem in \eqref{Problem1} as
\begin{eqnarray}
\label{Problem2}
    \mathcal{P}_2:\hspace*{2mm}&&\underset{\mathbf{Q}\succeq\mathbf{0},\bm{\Phi}}{\maxo}\hspace*{2mm}\overline{R}(\mathbf{Q},\bm{\Phi})\notag\\
    &&\hspace*{5mm}\mbox{s.t.}\hspace*{6mm}\mbox{C1:}\hspace*{1mm}\mathrm{Tr} (\mathbf{Q}) \leq n_t P_{T},\notag\\
    &&\hspace*{16mm}\mbox{C2:}\hspace*{1mm}\frac{\mathrm{Tr}(\mathbf{Q} \mathbf{T}_2)}{n_t}\mathrm{Tr}(\mathbf{R}_2(\mathbf{\Phi}\mathbf{\Phi}^H  - \mathbf{I})) + \sigma_d^2 \mathrm{Tr}(\mathbf{\Phi}\mathbf{\Phi}^H)\leq P_{A}.
\end{eqnarray}
% \begin{eqnarray}
% \label{Problem2}
%     \mathcal{P}_2:\hspace*{-4mm}&&\underset{\mathbf{Q}\succeq\mathbf{0},\bm{\Phi}}{\maxo}\hspace*{2mm}\overline{R}(\mathbf{Q},\bm{\Phi})\notag\\
%     &&\hspace*{5mm}\mbox{s.t.}\hspace*{7mm}\mbox{C1:}\hspace*{1mm}\mathrm{Tr} (\mathbf{Q}) \leq n_t P_{T},\notag\\
%     &&\hspace*{-1mm}\mbox{C2:}\hspace*{1mm}\frac{\mathrm{Tr}(\mathbf{Q} \mathbf{T}_2)}{n_t}\mathrm{Tr}(\mathbf{R}_2(\mathbf{\Phi}\mathbf{\Phi}^H  - \mathbf{I})) + \sigma_d^2 \mathrm{Tr}(\mathbf{\Phi}\mathbf{\Phi}^H)\leq P_{A}.
% \end{eqnarray}
Since the constraints are coupled and $\overline{R}$ is the difference between $\overline{R}_1$ and $\overline{R}_2$, $\mathcal{P}_2$ is non-convex. To this end, we propose an AO-based algorithm to optimize $\mathbf{Q}$ and $\mathbf{\Phi}$.
\subsection{Optimization of the Transmit Covariance Matrix}
For a given $\bm{\Phi}$, it can be shown that  $\overline{R}(\mathbf{Q},\bm{\Phi})$
is strictly concave with respect to $\mathbf{Q}$ \cite{Jun2020}. With the KKT condition, the original optimization problem $\mathcal{P}_2$ can be transformed into
\begin{eqnarray}
\label{Ori_Problem3}
    \mathcal{P}_3:&&\hspace*{-4mm}\underset{\mathbf{Q}\succeq\mathbf{0}}{\maxo}\hspace*{2mm}\mathcal{I}_1(\mathbf{Q})=\frac{1}{n_r}\log \det  (\mathbf{I} + c_1 c_2 \delta_1 \delta_2 \mathbf{Q} \mathbf{T}_2 )\notag\\
    &&\hspace{2mm}\mbox{s.t.}\hspace*{6mm} \mbox{C1:}\hspace*{1mm} \mathrm{Tr} (\mathbf{\mathbf{Q}}) \leq n_t P_{T},
    % &&\mbox{C2:}\hspace*{1mm} \mathrm{Tr} (\mathbf{\Phi} \mathbf{\Phi}^H) \leq n_l P_{A}
\end{eqnarray}
and solved using the water filling method. The optimal solution is given by
\begin{equation}
    \mathbf{Q}^{\mathrm{opt}} = \mathbf{U}_{\mathbf{T}_2} \mathbf{\Lambda_{\mathbf{Q}}} \mathbf{U}_{\mathbf{T}_2}^H, \label{waterfilling}
\end{equation}
where $\mathbf{U}_{\mathbf{T}_2} $ is a unitary matrix whose columns are all eigenvectors of $\mathbf{T}_2$, i.e., $\mathbf{T}_2 = \mathbf{U}_{\mathbf{T}_2}   \mathbf{\Lambda}_{\mathbf{T}_2} \mathbf{U}_{\mathbf{T}_2} $. The eigenvalues of $\mathbf{Q}$ is given by $\mathbf{\Lambda}_{\mathbf{Q}} = (\mu \mathbf{I} - (c_1 c_2 \delta_1 \delta_2 \mathbf{\Lambda}_{\mathbf{T}_2})^{-1} )^{+}$, where $\mu$ is the parameter chosen to satisfy the constraint $ \mathrm{Tr} (\mathbf{\mathbf{Q}}) \leq n_t P_{T}$.
%%%%%%%%%%%%%%%%%%%%%%%%%%%%%%%%%%%%
\subsection{Optimization the Reflection Matrix}
For a given $\mathbf{Q}$, the optimization of $\mathbf{\Phi}$ can be rewritten as
\begin{eqnarray}
\label{Ori_Problem4}
    \mathcal{P}_4:&&\hspace*{-6mm}\underset{\mathbf{\Theta},\mathbf{A}  \succeq 0}{\maxo}\hspace*{2mm} \mathcal{I}_2(\mathbf{\Theta}, \mathbf{A}) = \frac{1}{n_r}\Big(- \log \det(\mathbf{I} + \mathbf{F}_2 \mathbf{A}^2 ) + \log \det(\mathbf{I} + \mathbf{F}_1 e^{-\jmath\bm{\Theta}} \mathbf{A} \mathbf{T}_1 \mathbf{A}e^{\jmath\bm{\Theta}} )\Big) \notag\\&&\hspace{2mm}\mbox{s.t.}\hspace{2mm}\widehat{\mbox{C2}}\mbox{:}\hspace*{1mm}\frac{\mathrm{Tr}\big(\mathbf{Q} \mathbf{T}_2)}{n_t}\mathrm{Tr}(\mathbf{R}_2(\mathbf{A}^2 - \mathbf{I})) + \sigma_d^2 \mathrm{Tr}(\mathbf{A}^2)\leq P_{A}, \notag \\
    &&\hspace*{9mm}\mbox{C3:}\hspace*{1mm}\mathbf{A} =  \mathrm{diag} \{a_1, a_2, \cdots, a_{n_l}\},\hspace*{1mm}\forall i,\notag\\
    &&\hspace*{9mm} \mbox{C4:}\hspace*{1mm}\mathbf{\Theta} = \mathrm{diag} \{\theta_1, \theta_2,\cdots,\theta_{n_l} \}, \theta_i \in [0, 2 \pi),\hspace*{1mm}\forall i, 
    % &&\mbox{C2:}\hspace*{1mm} \mathrm{Tr} (\mathbf{\Phi} \mathbf{\Phi}^H) \leq n_l P_{A}
\end{eqnarray}
where matrices $\mathbf{F}_1$ and $\mathbf{F}_2$ are defined as $\mathbf{F}_1 = (c_1\delta_1 \sigma_d^2 \mathbf{I} + c_1 \delta_1 \delta_3 \mathbf{R}_2)$ and $\mathbf{F}_2 = c_1 \sigma_d^2 \alpha_1 \mathbf{T}_1 $, respectively. Moreover, we divide $\mathbf{\Phi} = \mathbf{A} e^{\jmath\bm{\Theta}} $ into two parts with $\mathbf{A} = \mathrm{diag} \{a_1, a_2, ... , a_{n_l}\}$ and $e^{\jmath\bm{\Theta}} = \mathrm{diag} \{e^{\jmath\theta_1}, e^{\jmath\theta_2}, ... e^{\jmath\theta_{n_l}} \}$. The optimization problem with respect to $\mathbf{\Phi}$ is non-convex due to the norm constraint of each element in $ e^{\jmath\bm{\Theta}}$ and the difference of two log determinants in the objective function. The gradients for the divided parameters are, respectively, given by
\begin{align}
\nabla_{\mathbf{\Theta}} \mathcal{I}_2 &=  \frac{2}{n_r}\Im\Big( \mathrm{diag} \left\{(\mathbf{I} + \mathbf{F}_1 \mathbf{M})^{-1}\right\} \Big) ,\label{grad_1} \\
\nabla_{\mathbf{A}}\mathcal{I}_2 &= \frac{2}{n_r} \Re\Big( \mathrm{diag} \left\{   (\mathbf{A} + \mathbf{F}_2\mathbf{A}^3)^{-1} \hspace*{-0.5mm}-\hspace*{-0.5mm} (\mathbf{A} + \mathbf{A} \mathbf{F}_1 \mathbf{M})^{-1}\right\}\hspace*{-0.5mm}\Big),\label{grad_2}
\end{align}
where $\mathbf{M}=e^{-\jmath\bm{\Theta}} \mathbf{A} \mathbf{T}_1 \mathbf{A}e^{\jmath\bm{\Theta}}$. We denote the total parameter matrix by $\bm{\Omega} = [\mathbf{\Theta}, \mathbf{A}]$, such that the gradient can be rewritten as $\nabla_{\bm{\Omega}} \mathcal{I}_2 = [\nabla_{\mathbf{\Theta}}, \nabla_{\mathbf{A}}]$. Here, we adopt the backtracking line search method \cite{boyd2004convex}. To start with, we define the constraint set of the optimization problem $\mathcal{P}_4$ as $\mathcal{Q}$. For a given $\widetilde{\mathbf{\Omega}}$, we denote the Euclid project operator by
\begin{equation}
\mathcal{P}_{\mathcal{Q}} (\widetilde{\mathbf{\Omega}}) = \underset{\mathbf{\Omega}\in \mathcal{Q}}{\arg\min }  ||\mathbf{\Omega} - \widetilde{\mathbf{\Omega}} ||_F.
\end{equation}
Assuming that the total parameter at the $t$-th iteration is $\bm{\Omega}^{(t)} = [\mathbf{\Theta}^{(t)}, \mathbf{A}^{(t)}]$, we search the best step size according to the following equation
\begin{align}
\gamma^* = \underset{\gamma \in [0, U]}{\arg\max } \, \mathcal{I}_2 (\mathcal{P}_{\mathcal{Q}} (\mathbf{\Omega}^{(t)} + \gamma \nabla_{\bm{\Omega}}\mathcal{I}_2)), \label{grad_search} 
\end{align}
where $U$ is a hyperparameter. Then, we update the phase transition and amplification matrix by  $\mathbf{\Omega}^{(t+1)} = \mathcal{P}_{\mathcal{Q}} (\mathbf{\Omega}^{(t)} + \gamma^* \nabla_{\bm{\Omega}}\mathcal{I}_2)$.
\par
The proposed algorithm is summarized in \textbf{Algorithm 1}. Note that for a given $\mathbf{\Phi}$, the solution $\mathbf{Q}$ of the $\mathcal{P}_3$ is optimal due to the concavity. On the other hand, for a given $\mathbf{Q}$, the adopted gradient line search method in $\mathcal{P}_4$ will converge because the objective function of $\mathcal{P}_4$ is monotonically increasing. Therefore, the proposed AO-based algorithm is guaranteed to converge.
\begin{algorithm}[t]
\caption{AO-Based Algorithm}
\begin{algorithmic}[1]
\small
\STATE Set iteration time $t=0$, convergence tolerance $0<\delta\ll1$ and  upper bound of search area $U$ and initialize the optimization variables $\mathbf{Q}^{(0)}$ and $\mathbf{\Omega}^{(0)} = [\mathbf{A}^{(0)}$, $\mathbf{\Theta}^{(0)}]$
\REPEAT
\STATE Calculate $\delta_i, i=1,\cdots, 4$, $\alpha_1$ and $\alpha_2$ according to \eqref{fixed_point_1}-\eqref{fixed_point_4}, \eqref{fixed_point_2_1} and \eqref{fixed_point_2_2} for given $\mathbf{Q}^{(t)}$ and $\mathbf{\Omega}^{(t)}$
\STATE Update the optimal $\mathbf{Q}^{(t+1)}$ according to \eqref{waterfilling}
\STATE Calculate the gradient $\nabla_{\bm{\Omega}} \mathcal{I}_2(\bm{\Omega}^{(t)})$ based on \eqref{grad_1} and \eqref{grad_2}
\STATE Obtain the optimal step size $\gamma^*$ by solving \eqref{grad_search}
\STATE Update the parameter $\mathbf{\Omega}^{(t+1)} = \mathcal{P}_{\mathcal{Q}} (\mathbf{\Omega}^{(t)} + \gamma^* \nabla_{\bm{\Omega}} \mathcal{I}_2)$
\STATE Set $t=t+1$
\UNTIL $\frac{\left|\overline{R}^{(t)}-\overline{R}^{(t-1)}\right|}{|\overline{R}^{(t)}|}\leq \delta$
\STATE \textbf{Output:}  $\mathbf{Q}^{(t)}$ and $\mathbf{\Omega}^{(t)}$
\end{algorithmic}
\end{algorithm}
%%%%%%%%%%%%%%%%%%%%%%%%%%%%%%%%%%%%%%%%%%
\section{Numerical Results}\label{S5}
In this section, we validate the accuracy of the DAs and the effectiveness of the proposed AO-based algorithm via numerical simulations. In the simulation, we adopt the following channel correlation matrix model \cite{ALMoustakas2003}
\begin{equation}
    \left[\mathbf{C}(\eta, \delta,  d_s)\right]_{m, n} = \int_{-180}^{180} \frac{\mathrm{d} \phi }{\sqrt{2 \pi \delta^2}} e^{2\pi\jmath d_s (m-n) \sin(\frac{\pi \phi}{180}) - \frac{(\phi - \eta)^2}{2 \delta^2}},
\end{equation}
where $m$ and $n$ denote the indexes of antennas or IRS elements, $d_s$ is the receive antenna spacing, $\phi$ represents the angular spread of the signal, $\eta$ is the mean angle, and $\delta$ denotes the root-mean-square angle spread. In the simulation, we set $\mathbf{T}_1 = \mathbf{C}(0, 30, 1)$, $\mathbf{T}_2 = \mathbf{C}(10, 5, 1)$, $\mathbf{R}_1 = \mathbf{C}(60, 30, 1)$, $\mathbf{R}_2 = \mathbf{C}(0, 30, 1)$. The path loss is set as $L_1 = L_2 = -25$ dB. The Monte-Carlo (MC) simulation results are illustrated by markers in all figures.
%%%%%%%%%%%%%%%%%%%%%%%%%%%%%%%%%%%%%%%%%%%%%%
\subsection{Accuracy of the DA}
%%%%%%%%%%%%%%%%%%%%%%%%%
\begin{figure}[t]
\centering 
    \centering
    \includegraphics[width=4.4in]{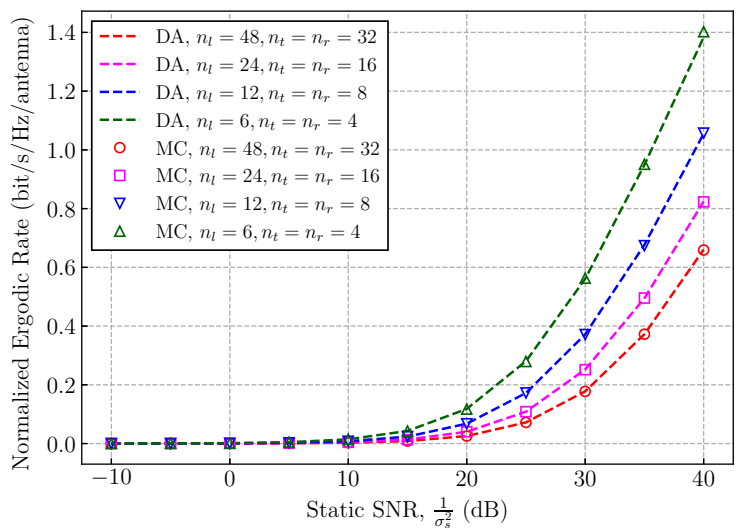}
\caption{The normalized ergodic rate versus signal to static noise ratio with $\sigma_d^2 = -30$ dBW. The dashed lines represent the DA analysis results.}
\label{Analysis_Conf_Ergrate_1}
\end{figure}
%%%%%%%%%%%%%%%%%%%%%%%%%%%%%%%%%%%%%%%%%
In Fig. \ref{Analysis_Conf_Ergrate_1}, we compare the results of the DA analysis with MC simulation. In the experiment, we set $\sigma_d^2 = -30$ dBW, the number of antennas as $n_t = n_r \in \{4, 8, 16, 32\}$, and the number of elements of the active IRS as $n_l \in \{6, 12, 24, 48\}$. The MC simulation results are obtained by $5000$ independent realizations of $\mathbf{X}_1$ and $\mathbf{X}_2$ in \eqref{rate}. As can be observed from Fig. \ref{Analysis_Conf_Ergrate_1}, the approximation is very accurate even in the small dimensional setting ($n_l = 4$).
%%%%%%%%%%%%%%%%%%%%%%%%%%%%%%%%%%%%%%%%%%%%%%
\subsection{Effectiveness of the Proposed Algorithm}
\begin{figure}[t]
\centering 
    \centering
    \includegraphics[width=4.4in]{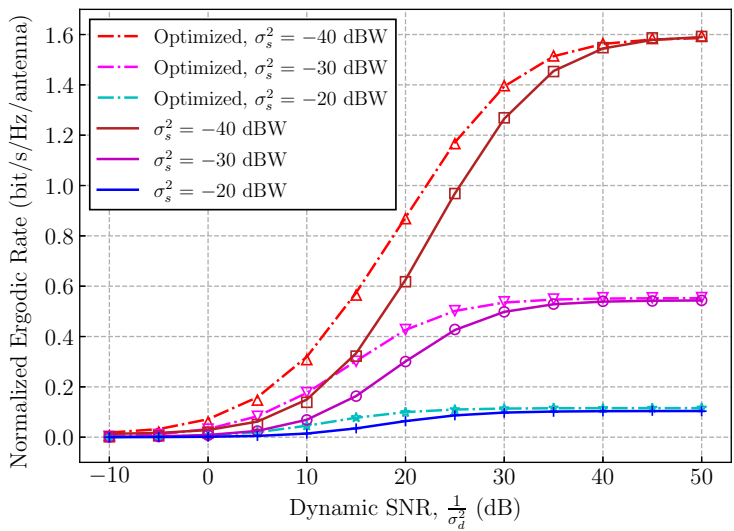}
\caption{The ergodic rate versus signal to dynamic noise ratio with $n_l = 18$, $n_t = n_r = 6$ and $P_T = P_A = 5$ W. The dash-dotted and solid lines represent the optimized and non-optimized results, respectively.}
\label{Opt_Conf_Ergrate_2}
\end{figure}
In Fig. \ref{Opt_Conf_Ergrate_2}, we show the effectiveness of the proposed algorithm with $n_l = 18$, $n_t = n_r = 6$ and $P_T = P_A = 5$ W. Comparing the solid and dash-dotted lines, we can observe that for a given $\sigma_s^2$, the proposed algorithm is more effective when $\sigma_d^2$ is not too large or too small. Moreover, as the dynamic noise decreases, the ergodic rate will first grow and finally saturate. This can be explained by the fact that, as $\sigma_d^2 \rightarrow  0$, the ergodic rate is monotonically increasing and upper bounded, cf. \eqref{rate}.
%%%%%%%%%%%%%%%%%%%%%%%%%%%%%%%%%%%%%%%%%%%%%%
\subsection{Active IRS versus Passive IRS}
% \begin{figure}[t]
% \centering 
%     \centering
%     \includegraphics[width=3.2in]{figure_2_1.eps}\label{Analysis_Conf_Opt_2}
% \vspace*{-2mm}
% \caption{The optimized ergodic rate versus signal to static noise ratio with $n_l = 18$, $n_t = n_r = 6$. The dashed lines and solid lines represent the optimized ergodic rate of active system and passive IRS system, respectively. The markers represent the Monte-Carlo results.}
% \label{Figure_2}
% \end{figure}
\begin{figure}[t]
\centering 
    \centering
    \includegraphics[width=4.4in]{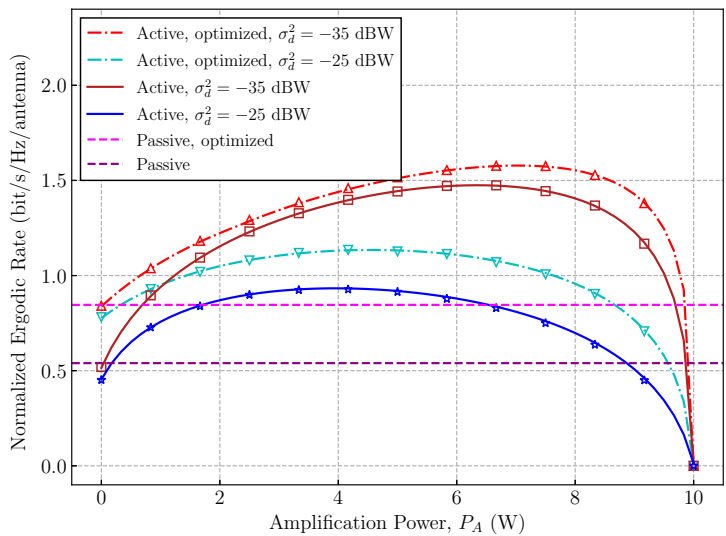}
\caption{The ergodic rate versus amplification power of the active IRS with $n_l = 12$, $n_t = n_r = 8$, $\sigma_s^2 = -40$ dBW and the total power $P_A + P_T = 10$ W. The dash-dotted and solid lines represent the optimized and non-optimized results of the active IRS, respectively. The dashed lines are optimized and non-optimized results of the passive IRS with $P_T = 10$ W.}
\label{Analysis_Conf_Compare}
\end{figure}

To show the gain introduced by the active IRS, we compare its normalized ergodic rate with that of the passive IRS in Fig. \ref{Analysis_Conf_Compare}. In the experiment, we set $n_l = 12$, $n_t = n_r = 8$ and $P_T + P_A = 10$ W for the active IRS and $P_T = 10$ W for the passive IRS. The static noises of two systems are set the same as $\sigma_s^2 = -40$ dBW. We have three observations from Fig. 4. First, as the power allocated to the active IRS increases, the optimized ergodic rate will increase at first and then decrease to $0$. Recall the fact that the transmit signal will suffer from the ``multiplicative fading" effect, but the signal amplified by the active IRS only goes through one hop, which avoids the severe path loss. As a result, when more power is allocated to the active IRS, the overall signal attenuation decreases, contributing to a rate improvement at the beginning. However, when the power of the transmitted signal is too small, the noise at the IRS will dominate the reflected signal, resulting in a significant performance degradation. Second, we can observe that the level of dynamic noise also influences the optimal power allocation policy. In fact, to increase the ergodic rate, more power should be allocated to the active IRS when the level of dynamic noise is not too high. Finally, even without optimization, the active IRS demonstrates its advantage over its passive counterpart. 
\section{Conclusion} \label{S6}
 In this paper, we investigated the benefit of using active IRSs in MIMO systems. For that purpose, we first derived the DA for the normalized ergodic rate, whose result was then utilized to jointly optimized the transmit covariance matrix and the reflection matrix, assuming only statistical CSI at the transmitter and the IRS. Numerical results validated the accuracy of the derived DA and the effectiveness of the proposed optimization algorithm. The analysis in this paper not only indicated that the active IRSs are a promising means to circumvent the ``multiplicative fading" effect, but also unveiled their consistent advantages over the passive ones. Besides, when the level of dynamic noise is relatively low, it is advisable to allocate more energy to the active IRS.
 %The analysis in this paper indicated that, the active IRSs can not only effectively circumvent the ``multiplicative fading" effect, but also demonstrate their consistent advantages over the passive ones.
%%%%%%%%%%%%%%%%%%%%%%%%%%%%%%%%%%
\section*{Appendix A \\ proof of Theorem 1}
To prove \textbf{Theorem. 1}, we first determine the DA of $\mathbb{E} [m_{\mathbf{B}_1}(z)]$ by the iterative method \cite{Xin2022}. Then, we show the existence and uniqueness of the DA using the contraction mapping and the normal family argument \cite{hachem2007deterministic}. 
\par
To derive the DA, we use the iterative method, i.e., take the conditional expectation and integrate out the randomness of $\mathbf{X}_i$ iteratively. Denote $\overline{\mathbf{T}}_1 = \widetilde{\mathbf{T}}_1^{\frac{1}{2}}(\widetilde{\mathbf{H}}_2 \widetilde{\mathbf{H}}_2^{H} + \sigma_d^2 \mathbf{I} )\widetilde{\mathbf{T}}_1^{\frac{1}{2}}$. Based on \cite[Theorem 1]{Couillet2011DE} with $K = 1$ and $\mathbf{S} = \mathbf{0}$ (due to the rotational invariance of Gaussian distribution, $\overline{\mathbf{T}}_1$ is not necessary diagonal), we can obtain the following relations
\begin{eqnarray}
\alpha \hspace{-2mm}&=&\hspace{-2mm} \frac{1}{n_l} \mathrm{Tr} [\mathbf{R}_1 (-z\mathbf{I} + \widetilde{\alpha} \mathbf{R}_1 )^{-1}],\\
\widetilde{\alpha} \hspace{-2mm}&=&\hspace{-2mm} \frac{1}{n_l} \mathrm{Tr} [\overline{\mathbf{T}}_1 (\mathbf{I} + \alpha \overline{\mathbf{T}}_1 )^{-1}], \\
m_{\mathbf{B}_1}(z|\mathbf{X}_2) \hspace{-2mm}&-&\hspace{-2mm} \frac{1}{n_r}  \mathrm{Tr} (-z\mathbf{I} + \widetilde{\alpha} \mathbf{R}_1 )^{-1} \xrightarrow[\overline{n} \rightarrow + \infty]{a.s.} 0, \label{DE_1} \label{as_convg}
\end{eqnarray}
where $m_{\mathbf{B}_1}(z|\mathbf{X}_2) = \mathbb{E}\big[ m_{\mathbf{B}_1}(z) |\mathbf{X}_2 \big]$ is the conditional expectation. We define the resolvent matrix of  $\overline{\mathbf{T}}_1$ by $\mathbf{\Upsilon}(z) = (-z\mathbf{I} + \overline{\mathbf{T}}_1)^{-1} = (-z\mathbf{I}  + \sigma_d^2 \widetilde{\mathbf{T}}_1 +  \widetilde{\mathbf{T}}_1^{\frac{1}{2}}\widetilde{\mathbf{H}}_2 \widetilde{\mathbf{H}}_2^{H} \widetilde{\mathbf{T}}_1^{\frac{1}{2}})^{-1}$ and denote $\widetilde{\mathbf{R}}_2 = \widetilde{\mathbf{T}}_1^{\frac{1}{2}} \mathbf{R}_2 \widetilde{\mathbf{T}}_1^{\frac{1}{2}}$. By referring to \cite[Theorem 1]{Couillet2011DE} with $K=1$ and $\mathbf{S} = \sigma_d^2 \widetilde{\mathbf{T}}_1$, we have
\begin{eqnarray}
\hspace{-4mm}\beta &\hspace{-2mm}=\hspace{-2mm}& \frac{1}{n_t}  \mathrm{Tr}  [\widetilde{\mathbf{R}}_2 (-z \mathbf{I} +\sigma_d^2 \widetilde{\mathbf{T}}_1 + \widetilde{\beta} \widetilde{\mathbf{R}}_2 )^{-1}], \\
\hspace{-4mm}\widetilde{\beta} &\hspace{-2mm}=\hspace{-2mm}& \frac{1}{n_t}  \mathrm{Tr} [ \widetilde{\mathbf{T}}_2 ( \mathbf{I} + \beta \widetilde{\mathbf{T}}_2  )^{-1}], \\
\hspace{-4mm}\mathbb{E} [m_{\overline{\mathbf{T}}_1}(z)] &\hspace{-2mm}-\hspace{-2mm}& \frac{1}{n_l}  \mathrm{Tr}  (-z \mathbf{I} + \sigma_d^2 \widetilde{\mathbf{T}}_1 + \widetilde{\beta} \widetilde{\mathbf{R}}_2 )^{-1} \xrightarrow{\overline{n} \rightarrow + \infty} 0. 
\end{eqnarray}
\par
Now we deal with $\mathbf{X}_2$ in \eqref{DE_1}. Further derivation shows that
\begin{eqnarray}
     % \widetilde{\alpha} &\hspace*{-2mm}=\hspace*{-2mm}& \frac{1}{n_l} \mathrm{Tr} [\overline{\mathbf{T}}_1 (\mathbf{I} + \alpha \overline{\mathbf{T}}_1 )^{-1}]\notag\\
     % &= \frac{1}{n_l} \mathrm{Tr} \big( (\mathbf{\alpha I})^{-1} -   (\frac{1}{\alpha} \mathbf{I} +  \overline{\mathbf{T}}_1 )^{-1} \big) \\
     \widetilde{\alpha} &\hspace*{-2mm}=\hspace*{-2mm}& \frac{1}{n_l \alpha} \mathrm{Tr} \big[ \mathbf{I} -  ( \mathbf{I} +\sigma_d^2 \alpha \widetilde{\mathbf{T}}_1 + \widetilde{\beta} \alpha \widetilde{\mathbf{R}}_2 )^{-1}\notag\\
     &\hspace*{-2mm}+\hspace*{-2mm}& (\mathbf{I} +\sigma_d^2\alpha \widetilde{\mathbf{T}}_1 + \widetilde{\beta} \alpha \widetilde{\mathbf{R}}_2 )^{-1} -  \frac{1}{\alpha}\mathbf{\Upsilon}(-\frac{1}{\alpha}) \big]\notag\\
     % &\hspace*{-2mm}=\hspace*{-2mm}& \sigma_d^2 \gamma_{\alpha} + \widetilde{\beta}(-\frac{1}{\alpha}) \beta(-\frac{1}{\alpha}) (\alpha c_2)^{-1} + o_{a.s.}(1), 
     &\hspace*{-2mm}=\hspace*{-2mm}& \overline{\alpha} + o_{a.s.}(1),
\end{eqnarray}
where $\overline{\alpha} =\sigma_d^2 \gamma_{\alpha} + \frac{1}{c_2 \alpha}\widetilde{\beta}(-\frac{1}{\alpha}) \beta(-\frac{1}{\alpha})$, $
    \gamma_{\alpha} = \frac{1}{n_l}  \mathrm{Tr} [\widetilde{\mathbf{T}}_1 (\mathbf{I} +\sigma_d^2 \alpha \widetilde{\mathbf{T}}_1 + \widetilde{\beta} \alpha \widetilde{\mathbf{R}}_2 )^{-1}]$ and $o_{a.s.}(1)
$ is a r.v. that almost surely converges to $0$. By denoting $[\delta_1, \delta_2, \delta_3, \delta_4] = [\frac{\alpha}{c_1}, \frac{\beta}{\alpha c_2}, \widetilde{\beta}, \gamma_{\alpha}]$, we can obtain \eqref{fixed_point_1}-\eqref{fixed_point_4}. By the almost sure convergence of \eqref{as_convg} and the dominated convergence theorem, we have
\begin{align}
    \mathbb{E} [m_{\mathbf{B}_1}(z|\mathbf{X}_2)] -  \frac{1}{n_r}\mathrm{Tr}(-z\mathbf{I} + \overline{\alpha} \mathbf{R}_1)^{-1} \xrightarrow{\overline{n} \rightarrow +\infty} 0,  
\end{align}
which proves \eqref{Th1_DE} in \textbf{Theorem. 1}.
\par
Next we will prove the existence of the system of equations. For notational simplicity, we define $\mathbf{\Psi}^{(0)}$ as $\mathbf{\Psi}^{(0)}=[\psi_1^{(0)}, \psi_2^{(0)}, \psi_3^{(0)}, \psi_4^{(0)}] = [\delta_1^{(0)}, \delta_1^{(0)}\delta_2^{(0)}, \delta_1^{(0)}\delta_3^{(0)},\delta_1^{(0)}\delta_4^{(0)}] = -\frac{1}{z}\mathbf{1}_4^T.$ Plugging $\mathbf{\Psi}^{(i)}$ in the system of equations, we can get $\mathbf{\Psi}^{(i+1)}$. Furthermore, we can prove by induction that
\begin{itemize}
    \item[i)]   $\psi^{(i)}_k, \overline{\psi}^{(i)} \in  \mathcal{S}(\mathbb{R}^+)$, where  $\overline{\psi}^{(i)} = - \frac{\psi_2^{(i)}\psi_3^{(i)}}{z({\psi_1^{(i)}})^2} - \frac{\sigma_d^2 \psi_4^{(i)}}{z\psi_1^{(i)}}$.
    \item[ii)]  for $z < 0$, $0 \leq \psi^{(i)}_k$, $\overline{\psi}^{(i)} \leq \frac{K}{|z|}$, where $K$ is a constant.
\end{itemize}
To prove that a function $\psi \in \mathcal{S}(\mathbb{R}^+)$, we need to validate that $\Im(\psi(z)) > 0$ for $z \in \mathbb{C}^{+}$, $\psi$ is analytic over $\mathbb{C}^{+}$ and $\lim_{y\rightarrow +\infty} - \jmath y \psi( \jmath y)$ converges\cite{hachem2007deterministic}. Due to the limited space, we only discuss $\psi_1^{(i)}$ here and the rest are similar. Let $\overline{\mathbf{Q}}^{(i)} = -\frac{1}{z}(\mathbf{I} + \overline{\psi}^{(i)} \mathbf{R}_1)^{-1}$, $\mathcal{E}^{(i)} = \max_{k\in\{1, 2, 3, 4\}} \{|\psi_k^{(i)} - \psi_k^{(i-1)}|, |\overline{\psi}^{(i)} - \overline{\psi}^{(i-1)}|\} $. For $z < 0$,  we have
\begin{eqnarray}
        |\psi_1^{(i+1)} - \psi_1^{(i)}| &\hspace*{-2mm}\leq\hspace*{-2mm}& |z||\overline{\psi}^{(i)} - \overline{\psi}^{(i-1)}|\notag\\
        &\hspace*{-2mm}\times \hspace*{-2mm}&
        |\frac{1}{n_r}\mathrm{Tr} (\mathbf{R}_1 \overline{\mathbf{Q}}^{(i)}\mathbf{R}_1\overline{\mathbf{Q}}^{(i-1)})| \notag\\
        &\hspace*{-2mm}\leq\hspace*{-2mm}& ||\mathbf{R}_1 ||^2 \frac{1}{|z|} \mathcal{E}^{(i)}=\frac{K_1}{|z|}\mathcal{E}^{(i)},\label{substraction}
\end{eqnarray}
where $K_1$ is a constant that is independent of $z$. Similarly, we have $\mathcal{E}^{(i + 1)} \leq \frac{K_{1,2,3,4}}{|z|} \mathcal{E}^{(i)}$. For $z < -C$ with $C > 0$ large enough, $ \psi_k^{(i)}(z)$ forms a  Cauchy sequence. So $\psi_k^{(i)}(z)$ has a unique limit denoted by $\psi_k(z)$ for $z < -C$. $\psi^{(i)}_k \in  \mathcal{S}(\mathbb{R}^+)$, so $\psi^{(i)}_k$ is analytic and uniformly bounded on each compact subset of $\mathbb{C} \backslash \mathbb{R}^{+}$. According to the normal family theorem\cite{rudin1987real}, $\psi_k$ is analytic on $\mathbb{C} \backslash \mathbb{R}^{+}$. Finally, $\psi_k \in \mathcal{S}(\mathbb{R}^+)$ can be proved by verifying that $\lim_{y\rightarrow +\infty} - \jmath y \psi_k(\jmath y) $ converges. The uniqueness of $\psi_k$  can be proved in the same way, i.e., assuming there exist two solutions $\psi_k$ and $\widetilde{\psi}_k$ both in $\mathcal{S}(\mathbb{R}^+)$, we can perform the subtraction like \eqref{substraction} to find the contraction. Therefore we complete the proof of \textbf{Theorem 1}.  \QED

\section*{Appendix B \\ proof of Theorem 3}
Denoting $t = \sigma_s^2$ in $\overline{R}_1$, we can get $\frac{\partial \overline{R}_1 }{\partial t} = \sum_i \frac{\partial \overline{R}_1 }{\partial \delta_i} \frac{\partial \delta_i}{\partial t} + \frac{1}{n_r}  \mathrm{Tr} ( t \mathbf{I} + (\delta_2 \delta_3 + \sigma_d^2 \delta_4) \mathbf{R}_1)^{-1} - \frac{1}{t}$. It can be shown that (due to space limitations, the proof is omitted here) $\delta_i$ is bounded for $t \geq \sigma_s^2$, i.e., $  \delta_i \in [c, C]$, where $c \geq 0$ and $C > 0$ are constants. So  $\lim_{t \rightarrow + \infty}\overline{R}_1(t) = 0$. Simple derivations show that $\frac{\partial \overline{R}_1 }{\partial\delta_i} = 0$. Thus $\frac{\partial \overline{R}_1(t) }{\partial t}$ is bounded and integrable for $t \in [\sigma_s^2, +\infty)$ and we have
\begin{eqnarray} 
  \overline{R}_1 &\hspace*{-2mm}=\hspace*{-2mm}& \int_{\sigma_s^2}^{ +\infty} - \frac{\partial \overline{R}_1}{\partial t} \mathrm{d}t \notag\\
        &\hspace*{-2mm}=\hspace*{-2mm}& \int_{\sigma_s^2}^{ +\infty}  \frac{1}{t} -  \frac{1}{n_r}  \mathrm{Tr} ( t \mathbf{I} + (\delta_2 \delta_3 + \sigma_d^2 \delta_4) \mathbf{R}_1)^{-1}  \mathrm{d}t \notag\\
        &\hspace*{-2mm}=\hspace*{-2mm}& \int_{\sigma_s^2}^{ +\infty}  \frac{1}{t} -  \mathbb{E} [m_{\mathbf{B}_1}(-t) ]\mathrm{d}t + o(1), 
\end{eqnarray} 
where $o(1)$ is a deterministic term that converges to zero. The same method can be applied to $\overline{R}_2$. Thus, we complete the proof. \QED

% Constraint $\overline{{\mbox{C1a}}}$ can be reformed to a LMI constraint by using general sign-definiteness \cite{zhou2020framework}. Constraint $\overline{{\mbox{C1a}}}$ can be reformed to another LMI constraint by S-procedure \cite{zhou2020framework}. 
% %\end{eqnarray}
% %The worst-case robust transmit power minimization problem is formulated as 
\bibliographystyle{IEEEtran}
\bibliography{Reference_List}
\end{document}